\documentclass{article}
\usepackage{latexsym}
\usepackage{graphicx}
\begin{document}

\title{Differentiation and Replication of Spots in
a Reaction Diffusion System with Many Chemicals}
\author{Hiroaki Takagi \and Kunihiko Kaneko \\
Department of Pure and Applied Sciences \\
University of Tokyo, Komaba, Meguro-ku, Tokyo 153, JAPAN}

\maketitle

\begin{abstract}
The replication and differentiation of spots in 
reaction diffusion equations are studied by extending the
Gray-Scott model with self-replicating spots to include
many degrees of freedom needed to model systems with many chemicals.  
By examining many possible reaction networks,
the behavior of this model is categorized into three types:
replication of homogeneous fixed spots,
replication of oscillatory spots, and differentiation from `multipotent spots'.
These multipotent spots either replicate or differentiate into other types of 
spots with different fixed-point dynamics, and as a result, an inhomogeneous 
pattern of spots is formed.
This differentiation process of spots is analyzed in terms of
the loss of chemical diversity and decrease of the local Kolmogorov-Sinai 
entropy.
The relevance of the results to developmental cell biology and
stem cells is also discussed.
\end{abstract}

A simple chemical reaction-diffusion system, the Gray-Scott 
(GS) model\cite{Gray&Scott}, has recently been studied extensively.
This model exhibits 
self-replicating spots within some range of parameter
values\cite{Pearson}.
Similar replicating spots have been found in experiment\cite{Lee}.  

In each model system studied to this time, only a single type of spots appear,
and there is no differentiation to create different types of spots.
To search for possibility in complex pattern formation as in biological systems,
it is interesting to study a reaction-diffusion system 
displaying not only the division but also differentiation
of such cell-like spots. There
it is more necessary to consider chemical reaction dynamics 
within each spot which are more complex than those adopted in the GS model.
Also, in order for spots to continue replication and differentiation,
their division process and their internal complex chemical dynamics
must somehow organize a proper relationship.
We consider a model with such sufficiently complex dynamics and study how this
relationship is organized. In so doing, we find the connection between 
self-replicating spots and cell differentiation in dynamical systems models
assuming the cell itself\cite{Kaneko&Yomo,Furusawa1,Furusawa2}. 
We do this by adopting a many-degree-of-freedom version of GS model, with 
the goal of realizing spot differentiation\cite{sentence}. 
%so as to extract some basic features of biological pattern formation 
%through reaction-diffusion scheme.

As a variant of the GS model we study the following model, possessing the 
many degrees of freedom needed to model a complex catalytic reaction network:
\begin{eqnarray*}
\frac{\partial u_i(x,t)}{\partial t}=D_u \nabla^{2} u_i(x,t)+A(1 - u_i(x,t))\\
- u_i(x,t)\sum_{j=1}^{Q}{\sum_{k=1}^{Q}W_i^{(j,k)}v_j(x,t)v_k(x,t)}\\
\frac{\partial v_i(x,t)}{\partial t}=D_v \nabla^{2} v_i(x,t)-Bv_i(x,t)\\
+ v_i(x,t)\sum_{j=1}^{P}{\sum_{k=1}^{Q}W_j^{(i,k)}u_j(x,t)v_k(x,t)}.
\end{eqnarray*}
Here $u_i(x,t)$ denotes the concentration of the $i$th inhibitor and 
$v_i(x,t)$ that of the activator chemical.  
Each inhibitor is produced at 
a constant rate, $A$, and both the activators and inhibitors
decay at rates proportional to their concentrations 
(with different coefficients).
The diffusion constants of inhibitors and activators are denoted by
$D_u$ and $D_v$, respectively. We assume that these constants are 
equal for all chemical species for simplicity.
This model is a variant of that proposed by Cronhjort and Blomberg 
\cite{Blomberg}, and is reduced to the GS model by setting $P=1$, $Q=1$, $W_0^{(0,0)}=1$. 
Here we fix $P=3$ and $Q=20$, and choose the reaction matrix $W$ randomly, under the constraint that the reaction for 
each activator's replication is catalized by $k$ (0$\le$$k$$\le$$K$) randomly chosen activators. (Here we have explored the cases in which $K=4,5,6$.) The specific numbers of chemicals $P$ and $Q$ are not important, but we note that it is very difficult to realize the differentiation we study, when these numbers are small.
While here we present results for the case of one spatial dimension, 
numerical results for the two-dimensional case indicate similar 
behavior.
   
The behavior of our model depends on the choice of the reaction network $W$ 
(which may be regarded to represent a prototype of a complex intracellular autocatalytic 
reaction network). We carried out numerical computations using a variety of 
randomly chosen reaction networks $W$, and with these we classified the possible types of
dynamics of our system.  Although we have studied a large number of networks,  
the observed behaviors belong to each of the four classes 
as will be shown.

For each set of simulations, this matrix was fixed, while the 
parameter values were set throughout as $D_u=1.0$, 
$D_v=0.010$, $A=0.020$, and $B=0.060$ or $0.070$.
 (Below we discuss bifurcation with the change of the parameters $A$ and $B$.)
These parameters were chosen so that the spot structure displayed by the GS 
equation is displayed by the presently considered model also.
We used $\Delta t=0.010$ and $\Delta x=1.0$ for the numerical integration, 
while we have confirmed that the numerical results are unchanged by using 
smaller values for $\Delta t$.

In general we begin from some initial conditions from which two simple spots 
are formed. These spots then produce additional spots, and the process 
continues until eventually spots are distributed throughout the entire system. 
When such a state is realized, spot division ceases. 
Here, we first classify the (transient) dynamics of our model starting 
from such initial conditions.
These results were obtained by considering several thousand randomly chosen 
reaction matrices.

To characterize the spot dynamics quantitatively, we have introduced
two quantities measuring the diversity of these dynamics. 
One is the chemical diversity $S_i(nT)$, which is defined by\\
$S_i(nT)=-\sum_{j=1}^{Q}P_i^j(nT)log(P_i^j(nT))$, \\
with $P_i^j(nT)=\overline{v_i^j}(nT)/\sum_{j=1}^{Q}\overline{v_i^j}(nT)$,\\
and $\overline{v_i^j}(nT)=(1/T)\sum_{t=(n-1)T}^{nT}v_i^j(t)$, ($n=$1,2,..),\\
where $v_i^j$ is the $j$th chemical concentration at the center of the $i$th 
spot.
The interval $T$ used for the average is chosen to be on the order of a time 
scale for spot division\cite{rem1}. 

The second quantity is the `local KS-entropy' $h_i$ of the $i$th spot. 
The local KS entropy here is defined as the sum of positive local Lyapunov exponents \cite{KK-PTP89}, 
obtained by using the tangent vectors corresponding to the chemical 
concentrations of the spot in question \cite{rem2}.

The spot dynamics are classified into the following four types.\\
{\bf (i) fixed-point case}\\
In this case, the set of concentrations within every spot converge to  
the same fixed point.  Although the spatial pattern itself is not completely 
homogeneous, each spot has identical,
fixed chemical concentrations, and they are separated by equal distance.\\
{\bf (ii) oscillatory case}\\
This case is further divided into non-chaotic, intermittent, 
and highly chaotic cases by the choice of reaction networks.
In the non-chaotic oscillation case, 
the spatial pattern of spots is fixed in time. In some reaction networks,
there exist propagating waves(See Fig.2a,2b).
In spatio-temporal intermittency,
the spatial pattern of spots is clearly separated into laminar and burst 
regions to form some characteristic patterns (See Fig.2c).  
Those patterns are expected to be in the same class that had been studied 
extensively in CML\cite{KK-PD89}.
Contrastingly, in the chaotic case,
the concentrations of each spot change chaotically in time, 
around the heteroclinic orbit(See Fig.2d). 
At any given time, the chemical concentrations vary from cell to cell,
but their averages over time are almost identical for each cell.
In this case, spots not only replicate but also sometimes annihilate. \\
{\bf (iii) case-I differentiation} \\
In this case, with time, spots differentiate into different types, namely, 
inner and outer types(See Fig.1a). For the inner type, 
the set of chemical concentrations converge to a fixed point, 
with less resource chemicals.
Outer type spots exhibit either fixed-point, periodic, or chaotic oscillations,
 depending on the reaction matrix and the parameter values.\\
Here, the diversity is larger for the outer type.  
The inner type has smaller chemical diversity and null local KS entropy,
as the dynamics fall onto fixed points.  For the parameter values we used,
the local KS entropy of the outer part can be either positive or zero,
depending on the network (See Fig.3a). (The sign also
changes depending on the parameter values.)\\
{\bf (iv) case-II differentiation} \\
Here, spots differentiate into two types, as in case I. Spots of the initial 
type exhibit chaotic oscillations. The division of these 
spots produces either the same type of spots with chaotic oscillations or 
different types with fixed-point dynamics.  
Here, differentiation is not governed by the inner/outer distinction,
but, rather spots with chaotic oscillations appear periodically, 
with some interval(See Fig.1b). Thus in this case 
there is pattern formation on two distinct spatial scales,
that of the spot size and that of the average distance between two 
chaotic spots.
In this case, the chaotic spots have a large chemical diversity, 
while other types have much smaller diversities.
Complex dynamics are stable within spots of the first type,
with positive local KS entropy, while for the other types, 
the local KS entropy is zero (See Fig.3b).

Let us study the differentiation process in more detail.
In both cases, spot types with smaller diversities and fixed-point dynamics
are differentiated from the initial type, with high diversity.
 
In fact, there exist initial conditions for which there results no 
differentiation of spots. Depending on the initial conditions and their 
chemical diversity, the first spot will either be a chaotic type or a 
fixed-point type. In the former case, there will exist differentiation, 
but in the latter case, there will be none, as such a spot can only replicate.
Thus, the differentiation from the first type is irreversible.  
For initial conditions with sufficiently large chemical diversity, a chaotic 
spot will appear. However, for smaller initial chemical diversities, the basin 
structure is so complex that it is not feasible to predict which type of spot 
will appear initially. If a chaotic spot does appear, the differentiation 
process we have been discussing will take place.

The difference between the two cases I and II is regard to the ability for
spontaneous differentiation.  In case I, spots located in the outer region are
maintained only through the flow of chemicals from the outmost part.
The differentiation from the outer region into the inner region 
continues if the dynamics in the outer region are chaotic.
However, if the dynamics of the outer spots converge to a fixed point, the 
differentiation is terminated, and as a result growth of the inner part ceases.

On the other hand, in case II,
the number ratio of the two types of spots remains nearly fixed (with small
fluctuations). In this case, differentiation from chaotic spots
continues with some rate, and the spot is distributed in 
space at some rate.
The pattern formed by the different types of spots is independent of the 
boundary conditions.

The differentiation here is reminiscent of that exhibited by stem cells 
in biological systems.
In a biological stem cell system, initial cell types can either replicate or 
differentiate into different types. These differentiated cells can themselves 
differentiate, or they can replicate. Upon such repeated differentiation,
multipotency comes to be lost, and eventually cell types that can only 
replicate are produced. 
In Ref.\cite{Furusawa2}, such stem cell is found to be natural characteristic 
of dynamical systems in which there exists cellular structures and 
cell division.
Here it is shown that such stem cell systems naturally appear in a 
reaction-diffusion system, without assuming the cellular structure in advance.

In case I of our model, these `stem-type' spots are located only 
in the outer region, while in case II,
such 'stem-type' spots exist within the inside of the pattern.
In both cases in which differentiation exhibits in our model,
the diversity and local KS entropy decrease
as the differentiation progress \cite{Furusawa2}.
Differentiation into a spot with fixed-point type dynamics is irreversible,
as mentioned above.
This is consistent with the loss of multipotency observed in real cellular
systems.

Finally, we examine the parameter dependence of 
each type of dynamics. Here we consider altering the parameters $A$ and $B$, 
while keeping the other parameter values fixed.
In the homogeneous case, the bifurcation that results is qualitatively 
similar to that for the original GS equation.  
Typical phase diagrams for differentiation in cases I and II are 
shown in Fig.4. As we see, as the decay rate for $v_i$ is decreased, there are bifurcations 
from a uniform system to a system with a single type of spots to a system with
turbulent spatial structure and no clearly defined spots. This set of 
bifurcations exists quite generally, for a variety of reaction networks $W$.

There has been a great deal of effort dedicated to relating reaction-diffusion 
systems to morphogenesis in biological systems, since the pioneering study of Turing \cite{Turing}.
However these models have not yet been able to describe the complexity 
observed in biological pattern formation, which includes cell differentiation 
from stem cells,
determination of fixed types, irreversibility and robustness in development.
In this work, we have found that the behavior which can be interpreted as
representing the irreversible differentiation from stem cells to fixed cell
types is displayed quite generally for a particular set of reaction diffusion
equations.
With regard to the broader context of the general study of reaction-diffusion
equations, it is interesting to note that
a pattern with two distinct spatial scales is organized in case II. Also it is
interesting to note that the behaviors from a variety of reaction networks
are classified just into the four cases. It will be
important to classify possible types of spatiotemporal patterns that appear 
when we increase the complexity of the internal reaction dynamics.

Since our model includes only reaction and diffusion, without any other 
type of mechanism, the present study is also relevant to the study of 
prebiotic evolution leading to proliferation and
diversification of cells.  Indeed, the autocatalytic nature of 
the reaction networks studied here is similar to that of the hypercycle 
studied by Eigen and Shuster \cite{M.Eigen}, while the relevance of the
spatial structure in resisting parasites\cite{Bresch} and short cuts of network
structure \cite{Niesert} has been discussed \cite{Blomberg,Hogeweg,McKaskill,Chacon} .
In contrast to previous studies, the present study explains not only 
the robust replication process but also the diversification to different cell types.
~\\

We would like to thank T.Yomo and C.Furusawa for stimulating discussions. 
This work is supported by Grants-in-Aid for Scientific Research from
the Ministry of Education, Science and Culture of Japan
(11CE2006, Komaba Complex Systems Life Project, and 11837004).
~\\

\begin{figure}
\begin{center}
\includegraphics[width=8cm,height=4cm]{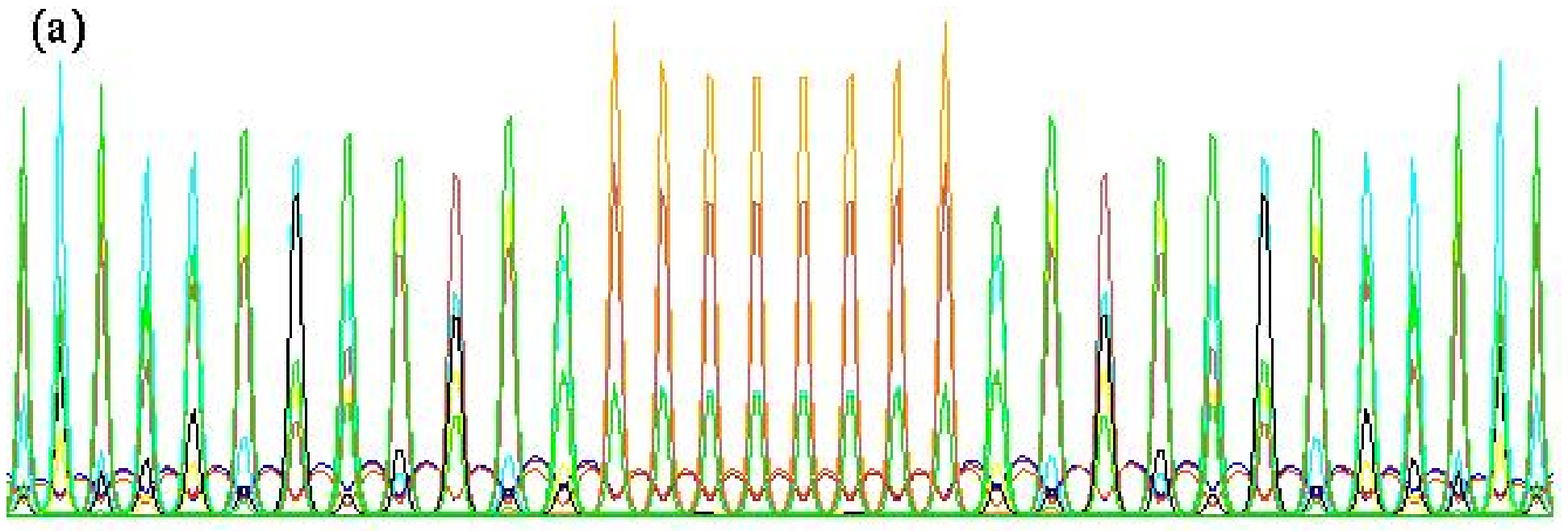}
\includegraphics[width=8cm,height=4cm]{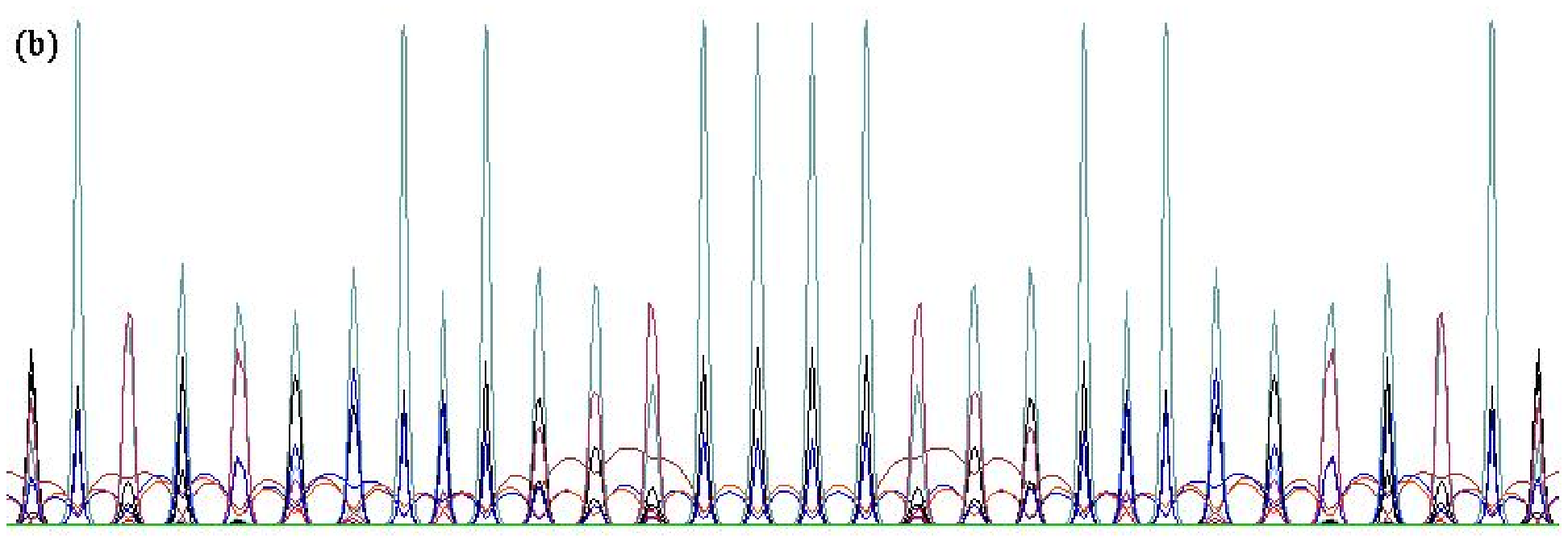}
\label{Fig.1}
\caption{Typical snapshot patterns of $u_i(x,t)$ and $v_j(x,t)$ are plotted.
Here we use a different color for each chemical species. At $t=0$ only a single spot with $v_j=0.250$ for all $j$ exists. Here (a) is case-I differentiation, and (b) case-II 
differentiation. (Only half of the pattern from the center to the right edge,
is plotted.)
%For the specific form of each reaction network $W$, 
%see http://chaos.c.u-tokyo.ac.jp/~takagi/research/GS, 
%although each of two cases appear for a wide variety of networks.
} 
\includegraphics[width=6cm,height=5cm]{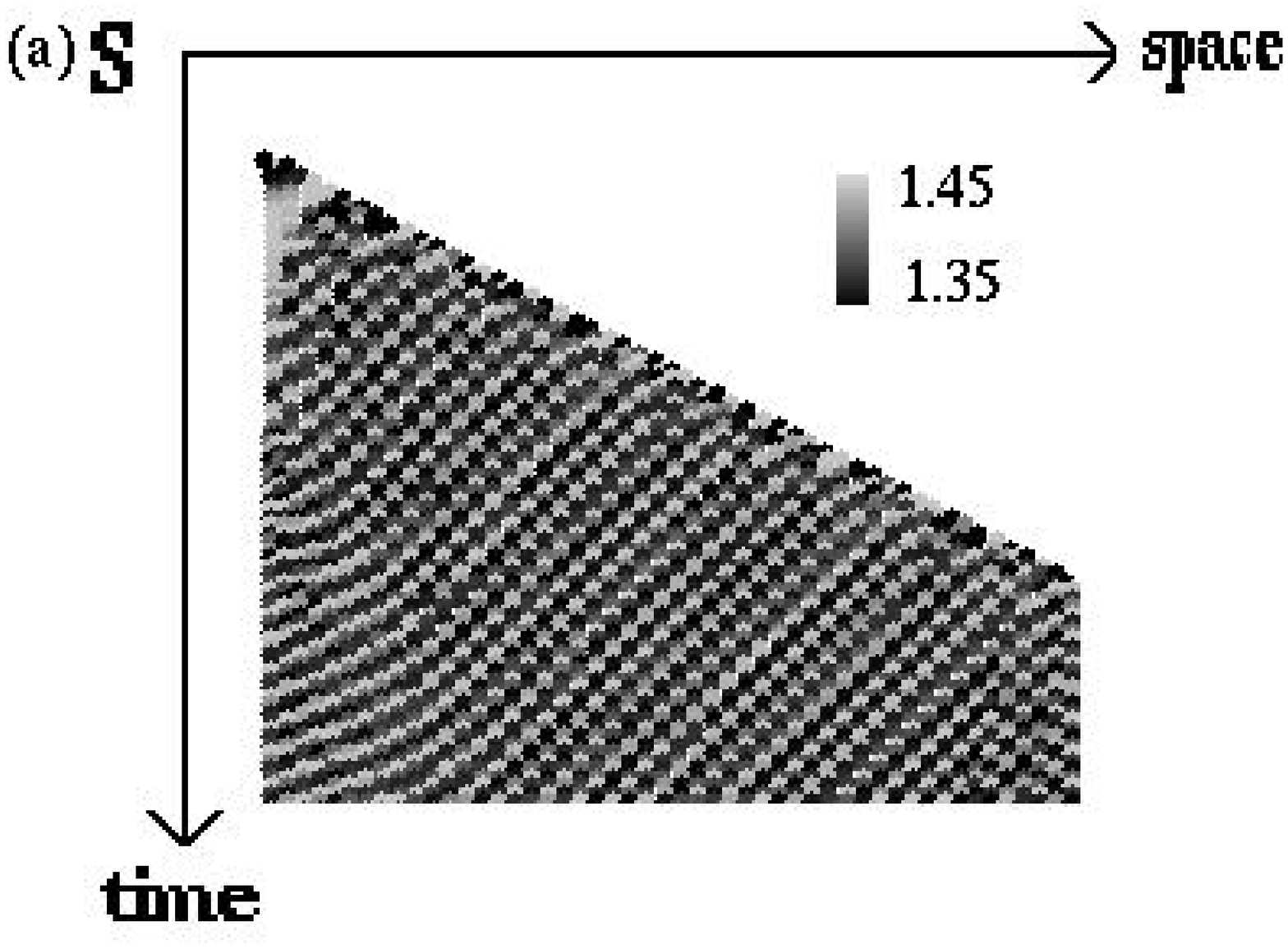}
\includegraphics[width=6cm,height=5cm]{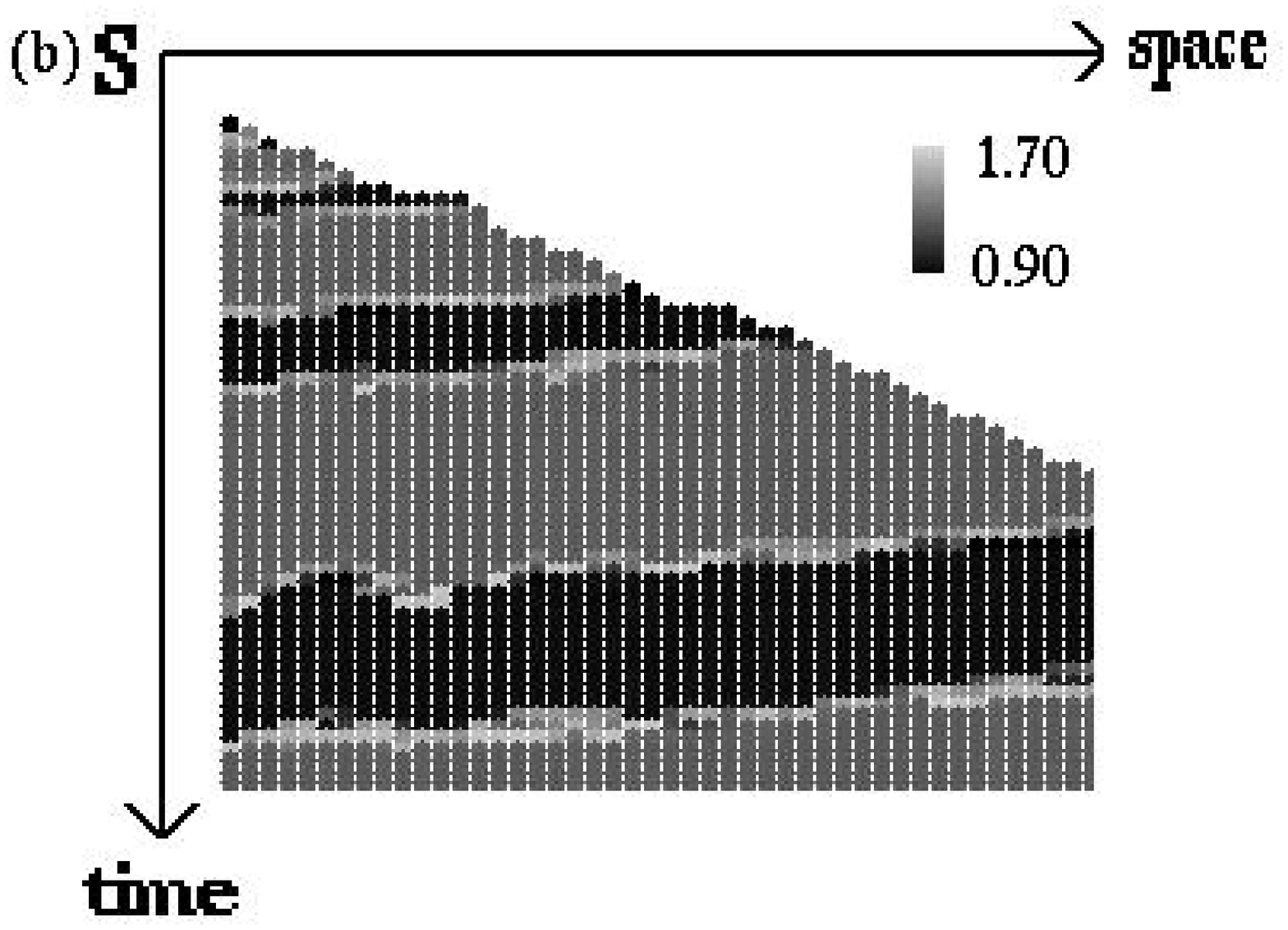}
\includegraphics[width=6cm,height=5cm]{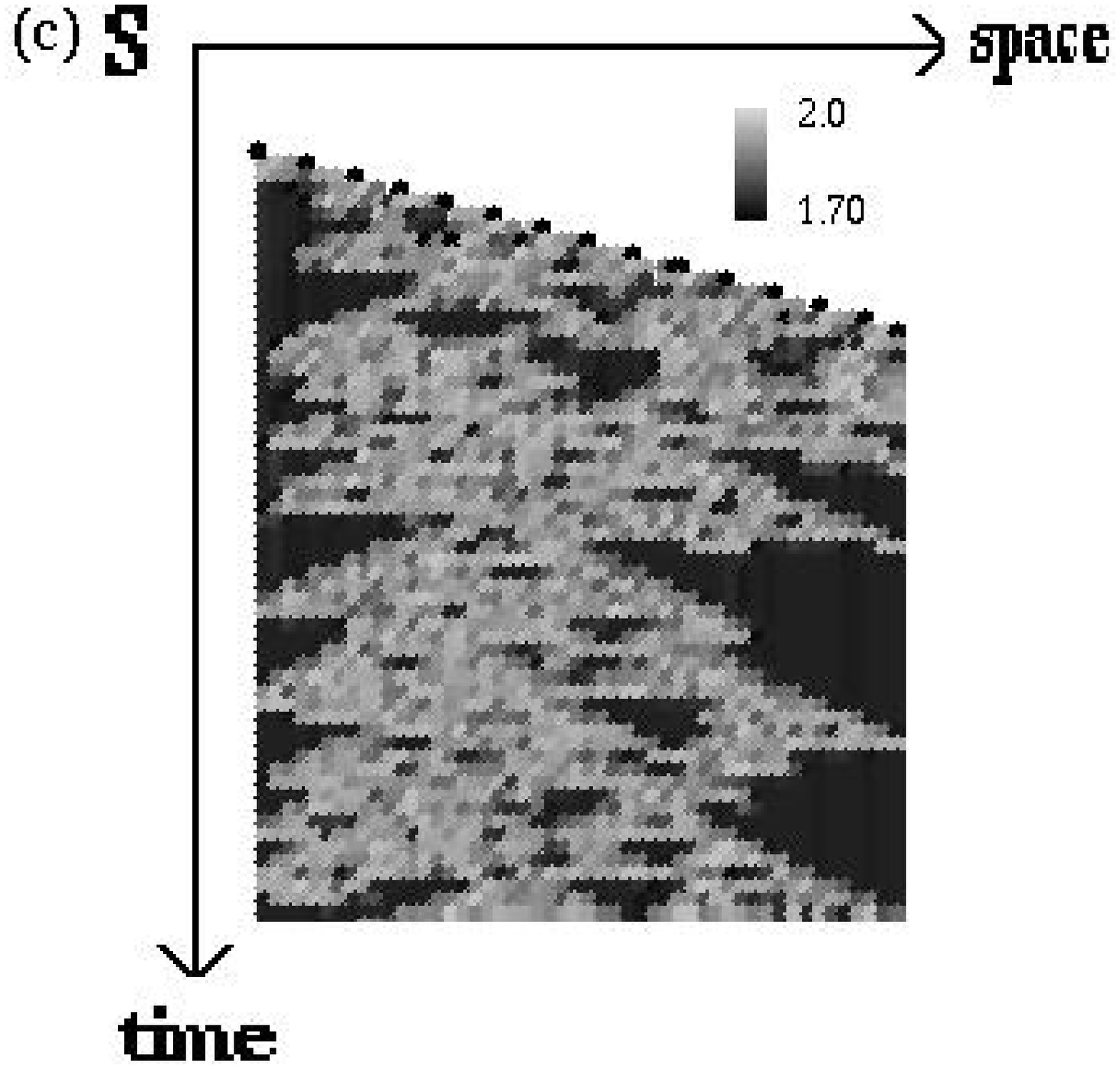}
\includegraphics[width=6cm,height=5cm]{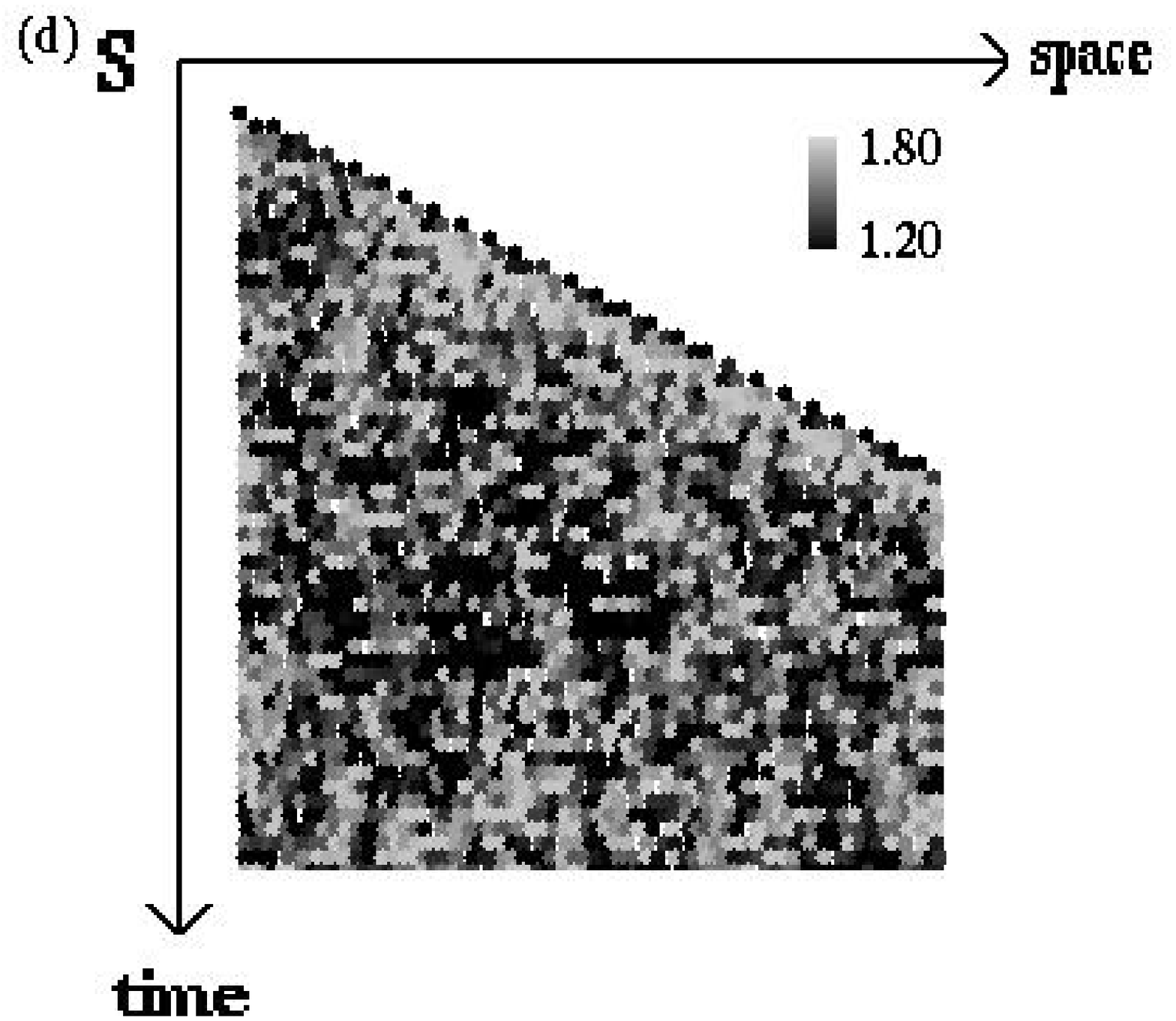}
\label{Fig.2}
\caption{Chemical diversity $S_i(nT)$ for some patterns, that are made by
different reaction networks. The diversity $S_i(nT)$ of the half-space 
(from the center to the right edge) pattern is plotted using a gray scale. 
(a) (b) the chemical dynamics of spots are non-chaotic;
(c) spatiotemporal intermittency is observed for chemical dynamics in
spots; (d) fully chaotic dynamics for spots.  For all these examples,
spot structures are preserved.}
\end{center}
\end{figure}

\begin{figure}
\begin{center}
\includegraphics[width=6cm,height=5cm]{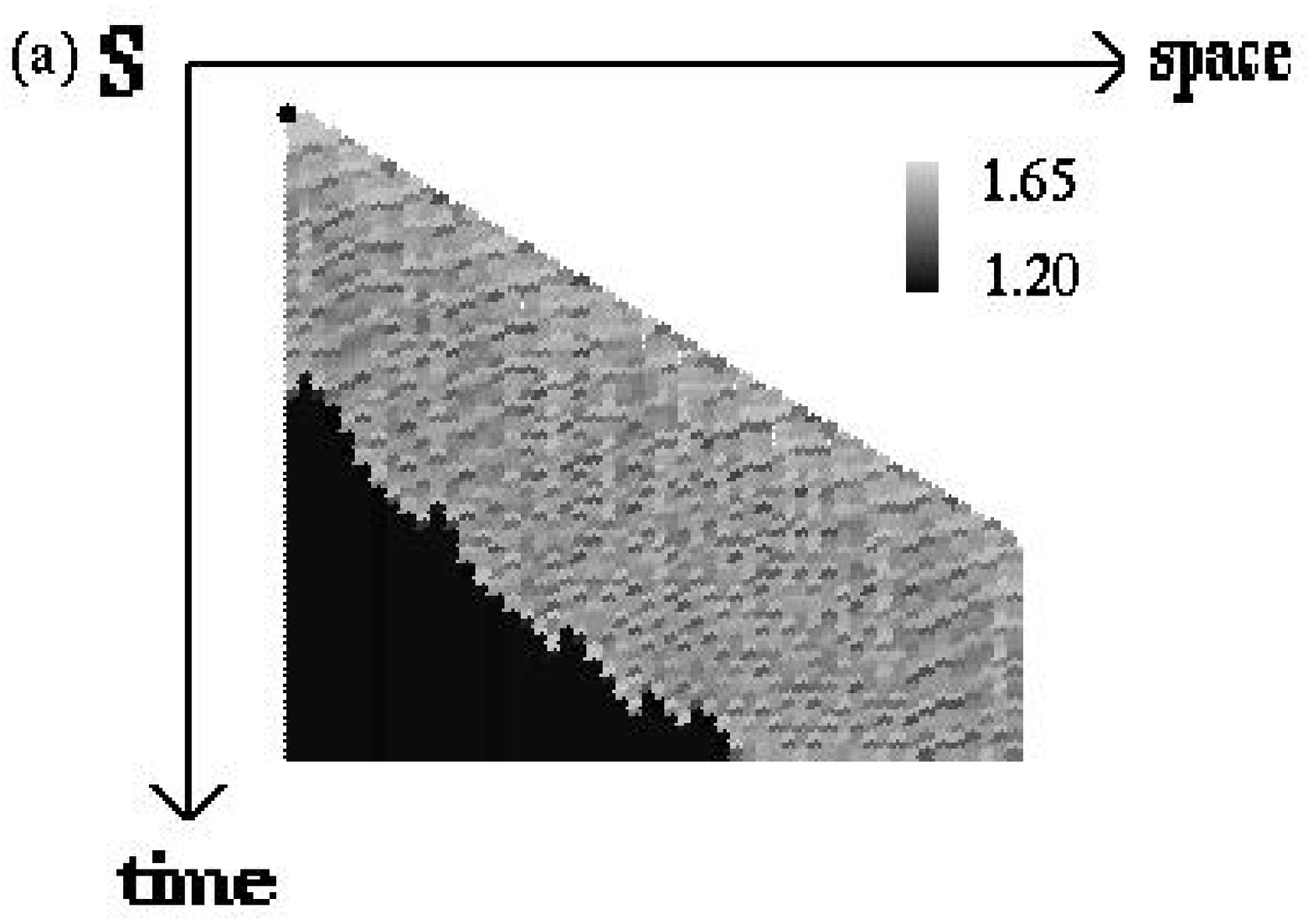}
\includegraphics[width=6cm,height=5cm]{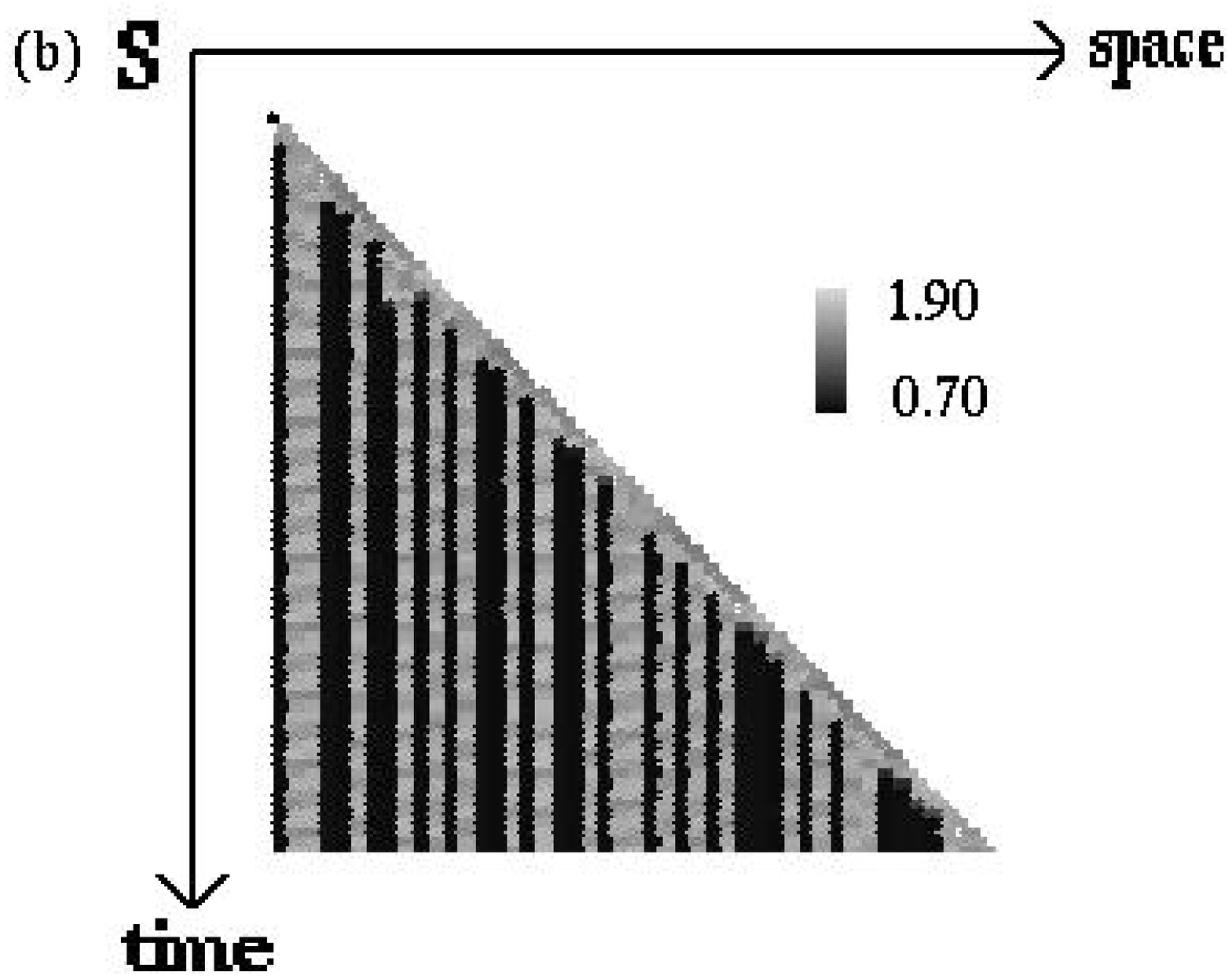}
\includegraphics[width=6cm,height=5cm]{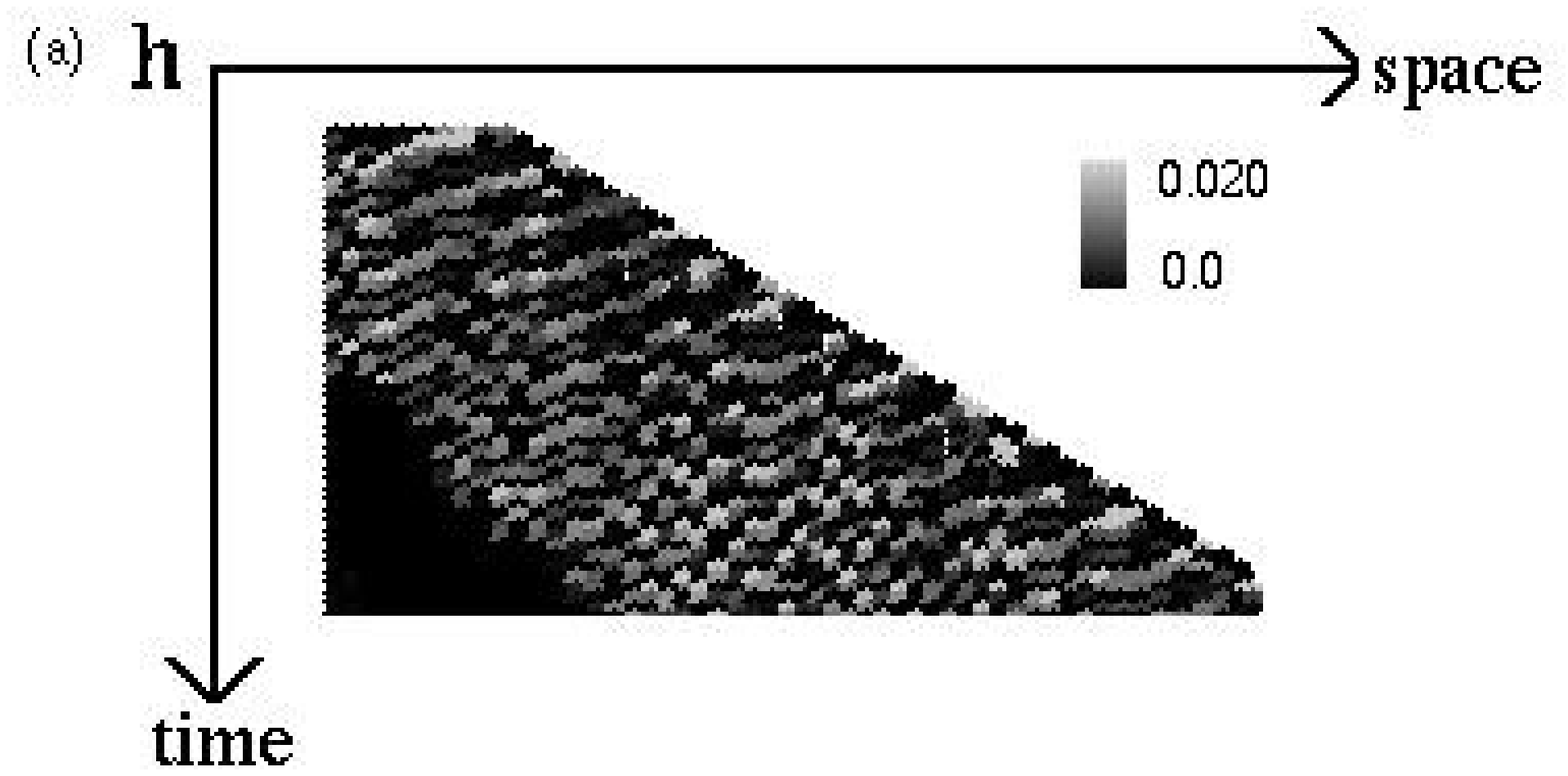}
\includegraphics[width=6cm,height=5cm]{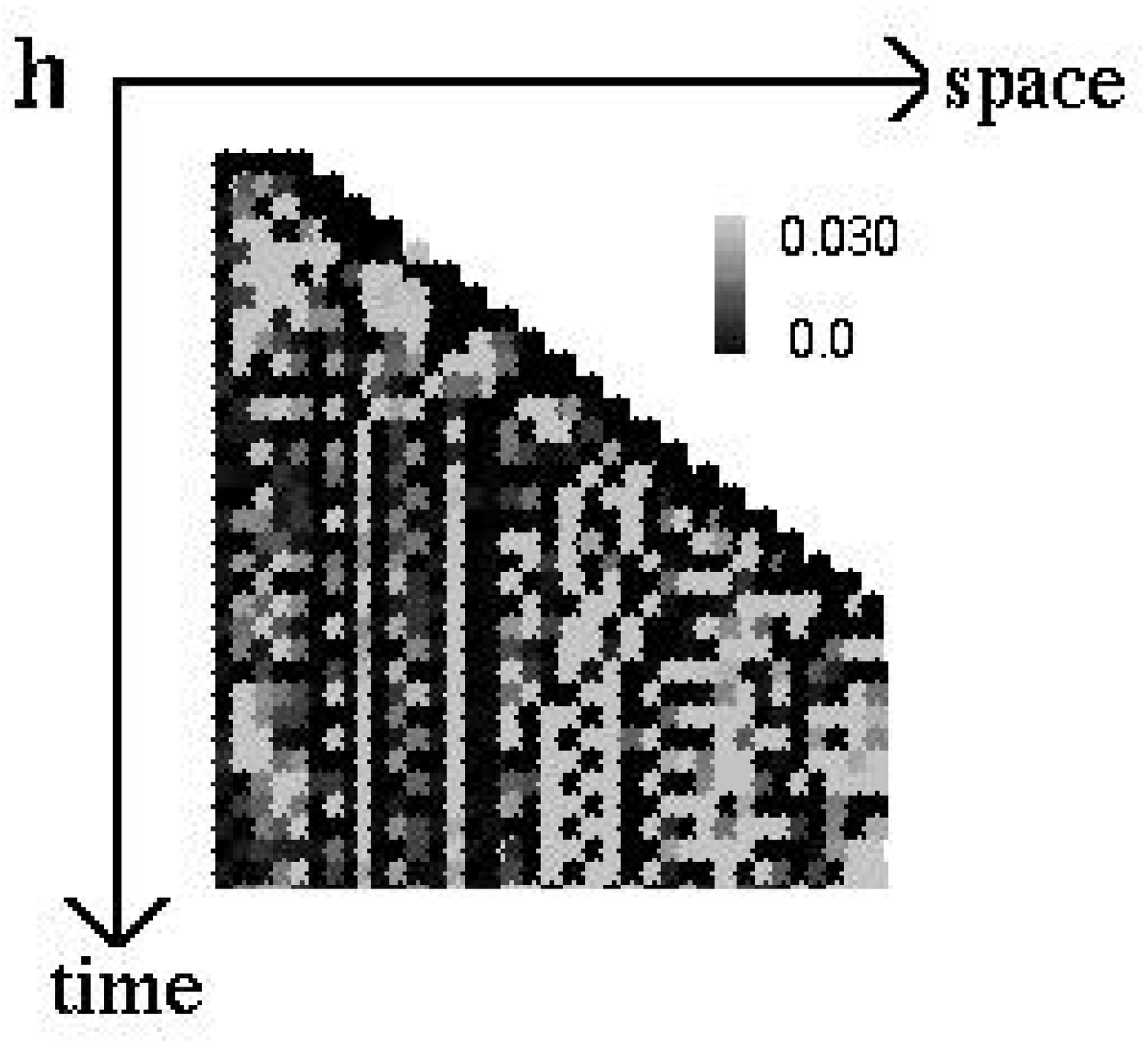}
\label{Fig.3}
\caption{Chemical diversity $S_i(nT)$ and local KS entropy $h_i$ for 
differentiation cases. The figures (a) and (b) are examples of
case-I differentiation while (c) and (d) are examples of
case-II differentiation. The same initial condition 
and reaction network as in Fig.1 were employed.}

\includegraphics[width=5cm,height=4cm]{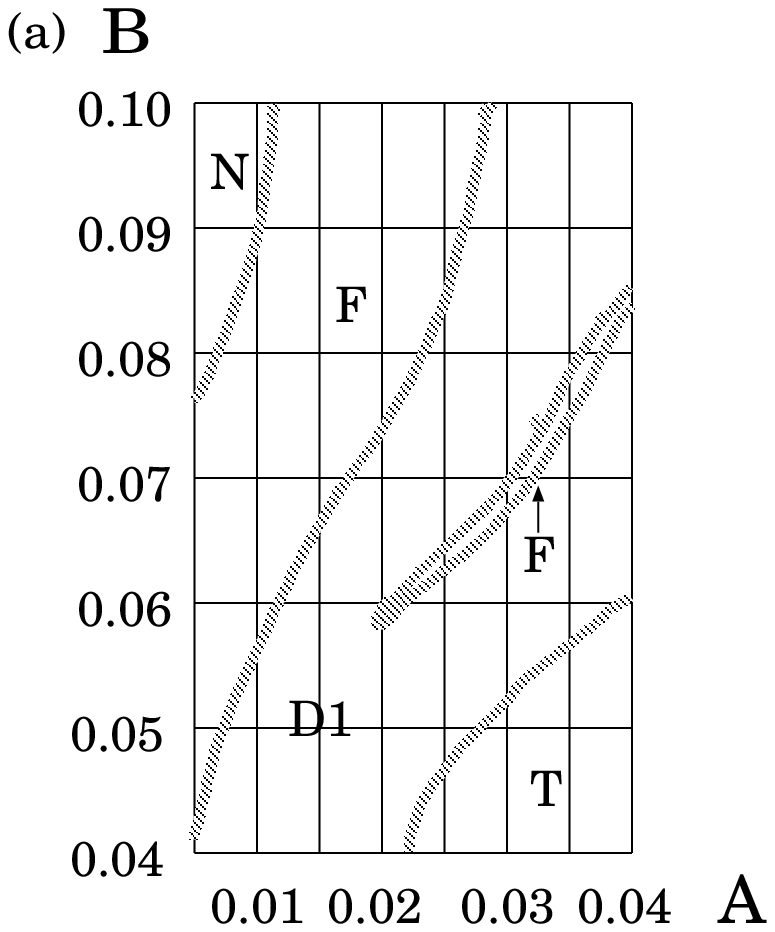}
\includegraphics[width=5cm,height=4cm]{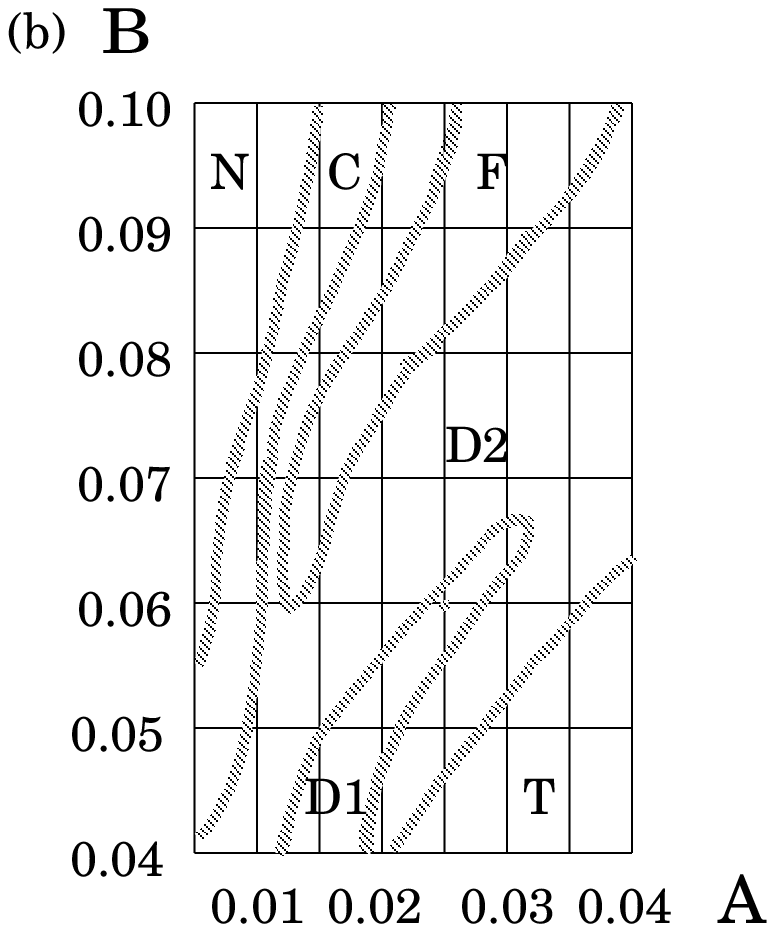}
\label{Fig.4}
\caption{Rough phase diagrams for (a) case I and (b) case II differentiation. 
The phases are denoted as follows:
$N$ corresponds to the spatially uniform state with $u_i=1$ and $v_j=0$ 
and no spots.
$F$ corresponds to fixed point case.
$C$ corresponds to chaotic case.
$D1$ corresponds to a case-I differentiation pattern.
$D2$ corresponds to a case-II differentiation pattern.
$T$ corresponds to turbulent chemical dynamics, with no clear spots.
(Initially, a single localized spot of a size 10, with $u_i=0.50$ and 
$v_j=0.250$ for all $i$ and $j$ was prepared.)}
\end{center}
\end{figure}


\begin{thebibliography}{99}
\bibitem{rem1}
Here, the interval $T$  was set to 1000, but the results are insensitive 
to this value as long as it is sufficiently large.
\bibitem{rem2}
More precisely, it is computed from the Jacobi matrix at each spot, 
averaging over time $T$, and neglecting diffusion out of the spot.
\end{thebibliography}
\end{document}